\documentclass[preprint,preprintnumbers,amsmath,amssymb,superscriptaddress]{revtex4-1}
\usepackage{fullpage,amsthm,amsmath,amsfonts,bm,times,color,bbm}
\usepackage{graphicx,subfigure}
\usepackage[english]{babel}
\usepackage{mathtools}
\usepackage{enumitem}
\usepackage{ccaption}
\usepackage{lineno}
\usepackage{float}
\def\degree{\hbox{$^\circ$}}

\def\eqa{\begin{eqnarray}}
\def\eea{\end{eqnarray}}
\newcommand{\eq}{\begin{equation}}
\newcommand{\ee}{\end{equation}}

\renewcommand{\>}{\rangle}

\newcommand{\el}{El Ni\~{n}o}
\makeatother

\makeatletter

\begin{document}

\title{
Network-based  Approach and Climate Change 
Benefits for Forecasting the Amount of Indian Monsoon Rainfall   
}
\author{Jingfang Fan}
\affiliation{Potsdam Institute for Climate Impact Research, 14412 Potsdam, Germany}
\author{Jun Meng}
\affiliation{Potsdam Institute for Climate Impact Research, 14412 Potsdam, Germany}
\author{Josef Ludescher}
\affiliation{Potsdam Institute for Climate Impact Research, 14412 Potsdam, Germany}
\author{Zhaoyuan Li}
\affiliation{School of Science and Engineering, The  Chinese University of Hong Kong (Shenzhen), 518172 Shenzhen, China}
\author{Elena Surovyatkina}
\affiliation{Potsdam Institute for Climate Impact Research, 14412 Potsdam, Germany}
\affiliation{Space Research Institute of Russian Academy of Sciences, Space Dynamics and Data Analysis Department, 117810 Moscow, Russian Federation}
\author{Xiaosong Chen}
\affiliation{School of Systems Science, Beijing Normal University, 100875 Beijing, China}
\author{J\"urgen Kurths}
\affiliation{Potsdam Institute for Climate Impact Research, 14412 Potsdam, Germany}
\affiliation{Department of Physics, Humboldt University, 10099 Berlin, Germany}
\author{Hans Joachim Schellnhuber}
\affiliation{Potsdam Institute for Climate Impact Research, 14412 Potsdam, Germany}

\begin{abstract}
The Indian summer monsoon rainfall (ISMR) has a decisive influence on India's agricultural output
and economy. Extreme deviations from the normal seasonal amount of rainfall can cause severe droughts or floods, affecting Indian food production and security.
 Despite the development of sophisticated statistical and dynamical climate models,
a long-term and reliable prediction of the ISMR has remained a challenging problem.  Towards achieving this goal, here we 
construct a series of dynamical and physical climate networks based on the global near surface air temperature field.
We uncover that
some characteristics of the directed and weighted climate networks can serve
as efficient 
long-term predictors 
for ISMR forecasting.
The developed prediction method produces a forecast skill of $0.5$ with a 5-month lead-time in advance by using the previous calendar year's data. The skill of our ISMR forecast, is comparable to the current state-of-the-art models, 
however, with quite a short (i.e., within one month) lead-time. We discuss the underlying mechanism of our predictor and associate it with network-delayed-ENSO and ENSO-monsoon connections. Moreover, our approach allows predicting the all India rainfall, as well as forecasting the different Indian homogeneous regions' rainfall, which is crucial for agriculture in India. We reveal that global warming affects the climate network by enhancing cross-equatorial teleconnections between Southwest Atlantic, Western part of the Indian Ocean, and North Asia-Pacific with significant impacts on the precipitation
in India. A stronger connection through the chain of the main atmospheric circulations patterns benefits the prediction of the amount of rainfall in India. We find a hotspots area in the mid-latitude South Atlantic, which is the basis for our predictor. Remarkably, the significant warming trend in this area yields an improvement of the prediction skill.
Based on our new forecasting framework, we predict the amount of rainfall of the future Indian summer monsoon in 2020 to be 842.85 mm, a -5\% departure compared to the long period average.




%

\end{abstract}
\date{\today}

\maketitle

\section{Introduction}
The All India Rainfall Index (AIRI) is the total amount of summer June-to-September (JJAS) rainfall averaged over the entire Indian
subcontinent. It is the Indian Meteorological Department's (IMD) primary indicator for monitoring the Indian summer monsoon rainfall (ISMR) \cite{rajeevan_new_2007}. The monsoon delivers more than 70 percent of the Indian annual rainfall. Extreme swings in the AIRI can cause severe economic and societal consequences \cite{rajeevan2001prediction,kumar_climate_2004,wahl_toward_2010}.
As a result, a long-term accurate prediction of the ISMR is crucial for taking timely actions  and mitigation activities. 
Over the last few decades, many efforts have been undertaken on (i) comprehensive diagnostic and statistical studies of climate and circulation 
\cite{shukla_empirical_1987,webster_monsoons:_1998,rajeevan_new_2007,di_capua_long-lead_2019} and (ii) atmospheric general circulation models (GCM) studies \cite{gadgil2011seasonal,rajeevan_evaluation_2012,ramu_indian_2016,delsole_climate_2012}, to improve the prediction skill of the ISMR.
%
%
However, both methods have the challenge of unstable relationships with the predictors over time as observed during the recent years \cite{rajeevan2001prediction,gadgil_monsoon_2005}. In particular, the
forecasting skill (cross correlation) of the official operational forecasts made by IMD is only  -0.12 for 1989--2012, the five
ENSEMBLE models' multi-model ensemble (MME) is 0.09 and the four APCC/CliPAS models's MME skill is 0.24 after 1989 \cite{wang_rethinking_2015}. 
It was argued that the recent failure is largely due to global warming \cite{wang_rethinking_2015}.  The maximum forecasting skill of $\sim 0.5$ was reported in  
some statistical and dynamical forecast models \cite{wang_rethinking_2015,delsole_climate_2012,delsole_artificial_2009}. However, the forecasting lead-time for these models is quite short, i.e., starting from May (one month lead-time) or June. Here we develop a statistical network approach with a forecast skill of around $0.5$, but at a significantly longer lead-time, 
by using the previous calendar year's data, i.e., 5-month lead time.
 
During the past decades, network theory has demonstrated its great potential as a versatile tool for exploring  dynamical and structural
properties of complex systems, from a wide variety of disciplines in physics, biology, and social science \cite{watts_collective_1998,barabasi_emergence_1999,newman2010networks,cohen2010complex,hens_spatiotemporal_2019}. It was widely used to predict the evolution of a scientist's impact \cite{sinatra_quantifying_2016}, forecast disease epidemics \cite{eubank_modelling_2004,brockmann_hidden_2013}, predict protein interactions and drug combinations \cite{kovacs_network-based_2019,cheng_network-based_2019}. Network science was also applied to uncover the institutional and corporate structure of the climate change counter-movement \cite{farrell_network_2016}.   Recently, it has been implemented in climate sciences
to construct Climate Networks (CN)~\cite{yamasaki_climate_2008,donges_complex_2009,feng_deep_2014,mheen_interaction_2013,fan2017network,boers_complex_2019}, in which the geographical locations are
regarded as network nodes, and the level of similarity between the climate records of these nodes represents the network links. The novelty of CN is that it maps out the topological features and patterns that are related to the physics of the dynamical climate variability.
Based on the CN approach, the forecasting capabilities were remarkably improved for the \el~Southern Oscillation \cite{ludescher_very_2014,meng_percolation_2017,meng_forecasting_2018}, extreme rainfall in the Eastern Central Andes \cite{boers_prediction_2014}, changes in atmospheric circulation under global warming \cite{fan2018climate} and onset/withdrawal of the Indian summer monsoon \cite{stolbova_tipping_2016}. Especially, the CN-based method proposed in \cite{stolbova_tipping_2016} allows predicting around 40 days in advance the Indian summer monsoon onset date, and 70 days in advance the withdrawal date. In the present work, we show that the CN approach fills a considerable gap between ISMR prediction skill and predictability. In particular, we forecasted that the AIRI for 2019 is 949.87 mm, which is less than 2\% deviation comparing to the observed 968.3 mm,
based on data until 2018. Additionally, we forecast a weaker and drier monsoon season with the AIRI 842.85 mm for 2020. We further analyze the consequences of global warming on climate networks structure  and  our network-based prediction skills.  


\section{Results}
\subsection{CN Construction}
\begin{figure}[H]
\begin{centering}
\includegraphics[width=0.95\linewidth]{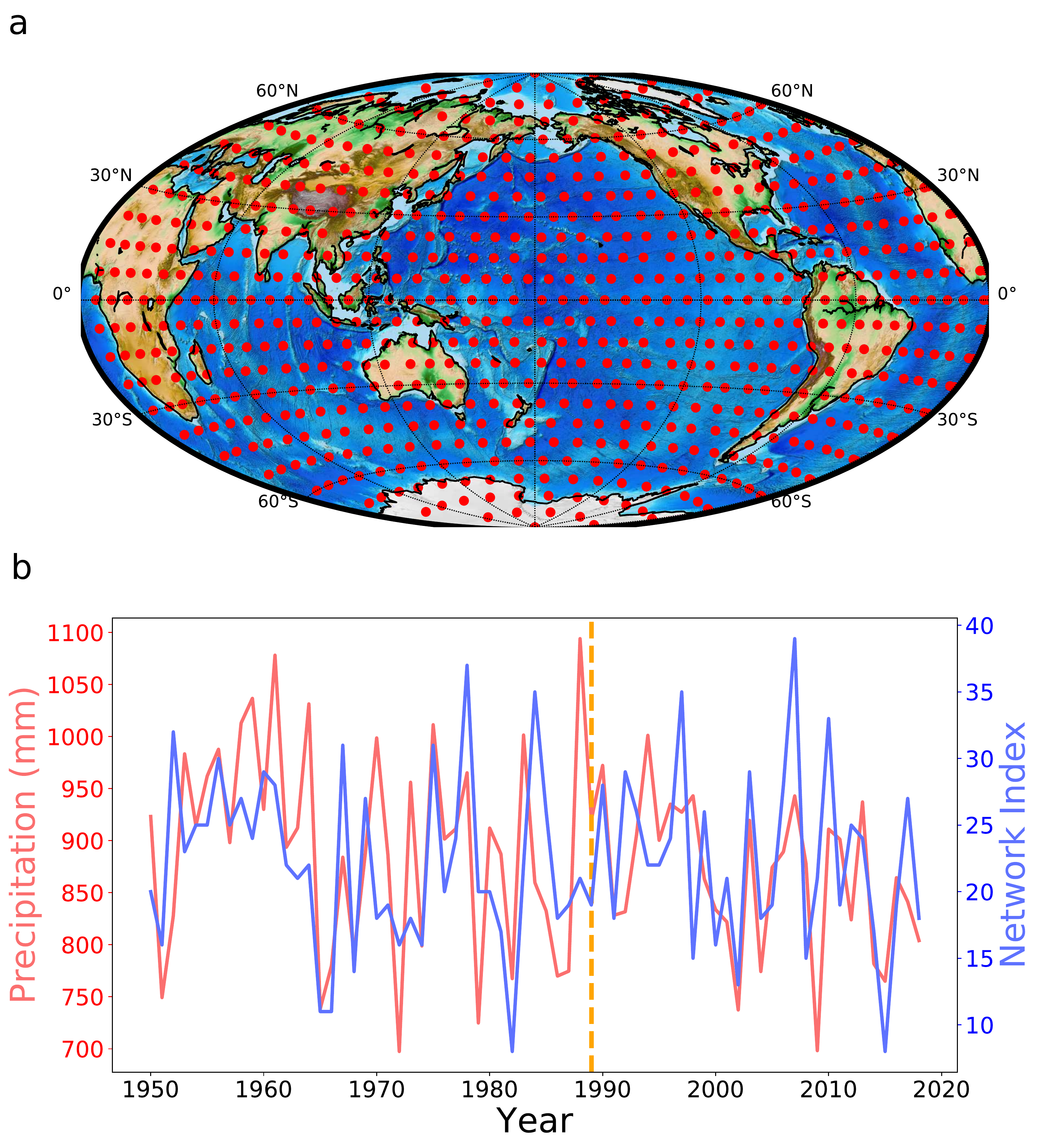}
\caption{\label{Fig1} 
{\bf Climate network, time series of observed AIRI and network prediction index.}
(a)  726 nodes (shown with small dots in red) that cover the globe approximately homogeneously. (b) The observed AIRI (units: mm, in red) and one network predictor index (in blue) during 1949-2018. The orange vertical  dashed line indicates the year of 1989.  The AIRI is the total amount of summer June-to-September rainfall averaged over the entire Indian subcontinent for each year, the network predictor index is calculated based on the previous calendar year's data.}
\end{centering}
\end{figure}

Based on the NCEP/NCAR reanalysis near surface daily air temperature anomalies data \cite{kalnay_ncepncar_1996}, we construct a CN for each calendar year since 1948 to the present.
 The 
 searching principle for this data field is, 
(1) it is  strongly correlated to the sea surface temperature (SST) and 
captures the
atmosphere--ocean--land interaction processes, and (2) it is  updated  timely so that we can  perform ISMR predictions regularly. In the current work, we choose 726 nodes (as shown in Fig.~\ref{Fig1}a), such that the globe is covered approximately homogeneously \cite{gozolchiani_emergence_2011}.
For each pair of nodes in the constructed CN, the link strength and direction are determined based on a similarity measure between the temperature anomaly time series of the nodes (for details in Methods). A pair of nodes is defined to be connected only if their link is within the top $5\%$ positive strength [see Eq.~(\ref{eq4})], which is also corresponding to a statistical significance of above $95\%$ confidence level [see Fig. S1].

In the following, we uncover some underlying dynamics of the global climate system related to the Indian summer monsoon by analyzing the variation of the corresponding CN properties over the years.  
%
%
Here, we focus on the \textit{degree} of the nodes, since it is one of the most fundamental parameters in network theory.
The degree of a node measures the total number of connected links \cite{albert_statistical_2002}. Since our CN is directed,  each node has two different degrees, the in-degree, which is the number of incoming  links (i.e., links that are coming from the other nodes); and the out-degree, which is the number of outgoing links (i.e., links that point toward the other nodes). Next, we construct yearly time series of the network in-degree and out-degree for each of the global nodes
based on the previous calendar year's data. Surprisingly, we uncover that strong and robust correlations exist between the observed AIRI and the in-degree time series for some key nodes on the globe. Based on these significant correlations,  we develop promising network predictors to forecast the ISMR with improved accuracy and
much longer lead-time compared to the state-of-the-art ISMR forecasting methods (see Fig. 1b for an example).

\subsection{Mining network predictors for AIRI}

\begin{figure}[H]
\begin{centering}
\includegraphics[width=0.85\linewidth]{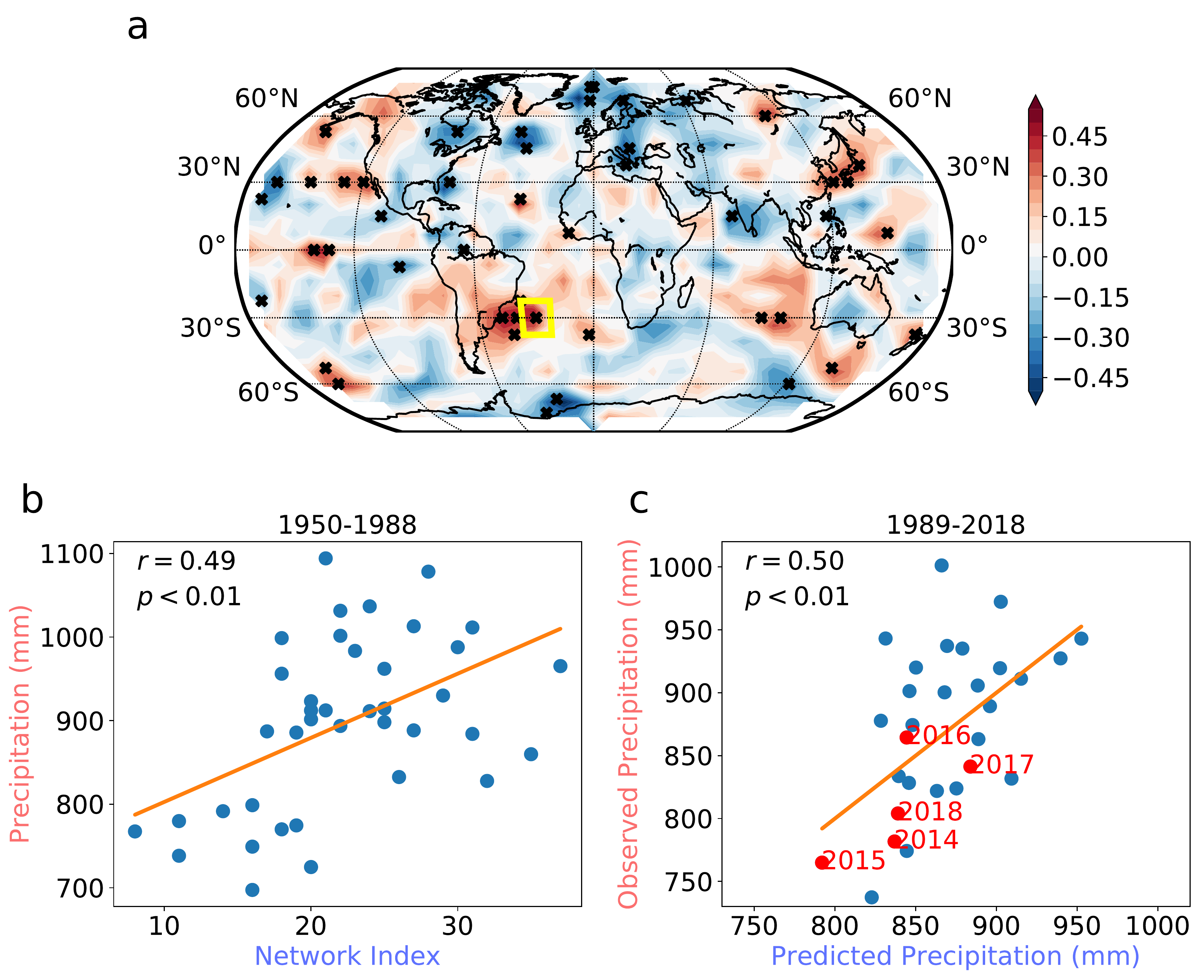}
\caption{\label{Fig2} 
{\bf AIRI--predictor correlations in the observations.}
(a)  The correlation coefficients between observed JJAS AIRI and in-degree time series during the training period for all nodes. (b) Scatter plots of the observed JJAS AIRI versus the  optimized network predictor (i.e., the in-degree time series of the key node in the yellow box) in the training period. (c) Scatter plots of the observed JJAS AIRI versus the predicted AIRI during the forecast period 1989-2018. Black $\times$ in  panel (a)
represent the regions with correlation significant at the 95\% confidence
level (Student's t-test). We highlight the recent five years scatter plots by red in panel (c).}
\end{centering}
\end{figure}

To \textit{mine} the dynamic network predictors for the AIRI, we calculate the cross correlation between the observed JJAS AIRI and the time series of network in-degree $K^{y}_{i}$ for each node $i$ on the globe. Note that in the calculation of the cross correlation we only compare each year's AIRI with its previous year's in-degree value, thus the correlation coefficient allows us to evaluate the prediction ability of each node's in-degree time series. 
Since the forecasting skill of the ISMR became very poor in the official operational forecasts and other models since the year 1989 \cite{wang_rethinking_2015}, 
we  perform the calculation on two consecutive periods, i.e., (i) 1950-1988,  as the training period; (ii) 1989-2018, the retrospective forecast period.
For the training period, the correlation coefficients are shown in Fig.~\ref{Fig2}a. As shown on the map, we find that the in-degree index for some nodes capture well the behavior of the AIRI. Particularly, we uncover that the node (in the region marked by yellow box in Fig. 2a) with the maximal correlation $r$ is located in the South Atlantic ocean [(30\degree S, 30\degree W)]. The value of this maximal correlation is $r=0.49$, with a student t-test significance of  $p<0.01$. Therefore, we choose the in-degree index of this key node as the optimized network predictor for the ISMR. In the following, we examine the forecasting capability of our network predictor during the forecast period 1989-2018.


To obtain the forecasted value of the AIRI for one specific year during 1989-2018, we substitute the forecasted year's predictor value (i.e., the previous year's network in-degree of the key node) into the least square linear regression equation. This equation used for prediction is derived using only the \textit{past} information, i.e., values of the AIRI and the network predictor for the period that between 1950 and the year before the forecasted year. Then we evaluate the practical predictability and reliability of our network predictor by calculating the cross correlation between the observed and forecasted AIRI values for the forecast period of 1989-2018. The independent forecast skill for
our forecast period 1989-2018 is obtained as $\sim0.50$, which is significantly higher than the IMD operational forecast skill [see Fig.~\ref{Fig2}c]. 
The result indicates the good accuracy of our network-based predictor with the long lead-time of 5 months.

Moreover, based on 2018's data, we forecasted that the AIRI for 2019 is  949.87 mm, ~7\% departure percentage compared to the long period average which is 887.5 mm. Note that the onset date of the 2019 Indian Summer Monsoon was significantly   
delayed, and the cumulative rainfall is even less than -30\% compared to the average at the beginning of July. However, after the beginning  of August, the precipitation has increased significantly, resulting in an actual AIRI of 968.3 mm.
Taking into account that the magnitude of the AIRI could have reached more than 20\%,  
we find that our forecasted AIRI value for 2019 is very close to the observed one [deviation less than 2\%].

\subsection{Key Node Features}

We identify as the key node that has the maximal correlation between its in-degree time series and the AIRI. It is located in the South Atlantic Ocean (30\degree S, 30\degree W). 
From a meteorological perspective, the teleconnection between the key node's location in mid-latitude of the South West Atlantic Ocean and the Indian subcontinent is no coincidence. The orography of the Eastern coast of the South American continent is prone to the formation of an anticyclone at mid-latitude due to the interaction with mid-latitude Westerlies (winds that blow from the west at the surface level between 30\degree and 60\degree S).
Moreover, there is the influence of the South polar jet stream, which appears within the upper air Westerlies. The acceleration/deceleration of the South polar jet stream induces areas of low/high pressure respectively, which link to the formation of cyclones and anticyclones.  Because the South polar jet stream strengthens and weakens seasonally, and from year to year, it synchronizes with the variability in the strength of the circulation centered around the location of the key node.

A critical question, then, is how the anticyclonic circulation in South Atlantic connects with the amount of rainfall of the Indian Summer monsoon. One of the possible explanations  might be the following: the circulation around the key node in mid-latitude Westerlies appears in the meander of the south polar jet stream, which connects to the anticyclone around the Mascarenas High near Madagascar. Meanwhile, Mascarenas High, in turn, connects to the very fast Somali jet, which greatly enhances southwest monsoon wind bringing rainfall to the Indian subcontinent. This strong cross-equatorial flow from the southern to the northern hemisphere impacts significantly on the amount of rainfall during the Indian summer monsoon. One of the possible  data-based teleconnection path is shown in Fig. 3. Such, the key node location is sort of the nexus of the South Atlantic circulation, the south polar jet stream and the main atmospheric circulations patterns over the Indian Ocean. 
It explains why we observe a significant correlation between 
the network characteristics of the key node and the AIRI.


\begin{figure}[H]
\begin{centering}
\includegraphics[width=1.0\linewidth]{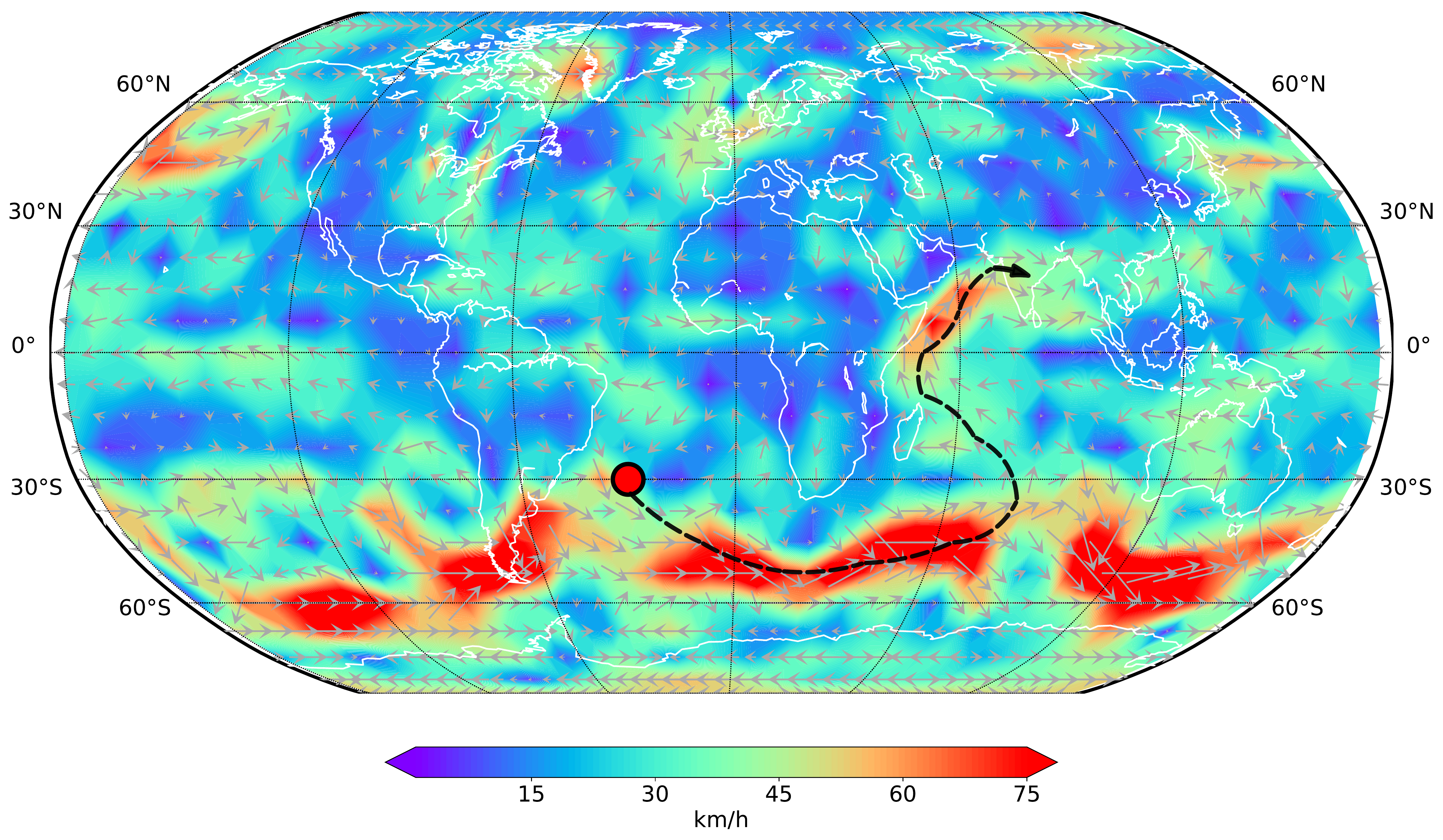}
\caption{\label{Fig3_a} 
{\bf Teleconnection path between the key node in South Atlantic and Indian subcontinent.} The red point indicates the key node, and the dashed black curve is a possible teleconnection path driven by wind patterns. The anticyclonic circulation around the key node in mid-latitude Westerlies appears in the meander of the south polar jet stream, which connects to the anticyclone around Mascarenas High near Madagascar that, in turn, connects to the Somali jet. This high-speed jet greatly enhances the southwest monsoon wind, which brings rainfall to the Indian subcontinent. The colors and grey arrows depict the magnitudes and directions of the 850 hPa winds. Here we show a typical  example of the observed wind data on 12th August 2014 during the Indian summer monsoon season.
}
\end{centering}
\end{figure}

\subsection{Potential Physical Mechanism}
\subsubsection*{Network Delayed ENSO Teleconnection}
To reveal the physical mechanism why our network predictor performs so well, we first analyze the network structure for some special years as typical examples. In Fig.~S2, we present all the in-direction links toward to the key node (in the yellow box of Fig. 2a) for 4 consecutive years 1980-1983. Interestingly, we find that the number of the in-direction links (in-degree) reaches a minimum in 1981, thus predicts a weak monsoon season for the following year 1982 (see Fig. 1b), which is an El Ni\~{n}o onset year. 
%
Fig.~S3 and S4 show another two analogous examples for different periods that also include the 1991 and 2002 El Ni\~{n}o onset years. All these examples explicit 
the network delayed ENSO teleconnection, i.e., the in-degree index is locally minimized in the previous calendar year of the El Ni\~{n}o onset. Further evidence for the robustness of this teleconnection is demonstrated in Fig. S5. 
We uncover that the network predictors are located in local valleys in  17 out of the 20 El Ni\~{n}o events which occurred during the period of 1950-2018, with only 3 exceptions, including, 1997, 2006 and 2009. Moreover, in Fig. 4(a) we find that our network predictor is inversely correlated with the following year's JJAS Ni\~{n}o 3.4 SST anomaly, and the correlation is $r=-0.43$ with $p<0.01$. This shows that the network response to ENSO is well captured in our framework.
%

To analyze the physical reason for the lack of in-directed links to the key node during the previous calendar years of El Ni\~{n}o onsets, here we provide a possible explanation for links in our CN as teleconnections driven by the planetary waves--(Rossby wave)~\cite{boers_complex_2019,wang_dominant_2013,zhou_teleconnection_2015}, which could circulate and propagate either intra or inter hemispheres \cite{li_equatorial_2019,boers_complex_2019}. This can be used to explain the link structure in our CN, e.g., as shown in Figs.~S2, S3 and S4. Where we have both cross-equatorial and intra-Hemisphere links.
%
In addition, we find that the key node [in the yellow box of Fig.~\ref{Fig2}a] is located very close to the main cross-equatorial wave ducts  for both boreal winter  and summer seasons \cite{li_equatorial_2019}, which reveals  important (special) physical and climatological implications of the key node's geographical location. It has been reported that during the previous calendar year of an El Ni\~{n}o onset year, the temperature variations in the Ni\~{n}o 3.4 region where the El Ni\~{n}o phenomenon originates become more disordered \cite{meng_forecasting_2018,meng_percolation_2017,meng2019complexity}. As a consequence the similarity (correlation) between the temperature variations in different locations of the equatorial region, particularly, in the tropical Pacific Ocean, are weakened or even destroyed.
Therefore we suspect that the high disorder phenomenon in the equatorial region happening before the El Ni\~{n}o onset might form a barrier that destroys equatorial links and also blocks tele-links across the equator. Our results shown in Figs.~S2-S5 strongly verify this hypothesis. 
%


%

\subsubsection*{ENSO-Monsoon teleconnection}
\begin{figure}[]
\begin{centering}
\includegraphics[width=1.0\linewidth]{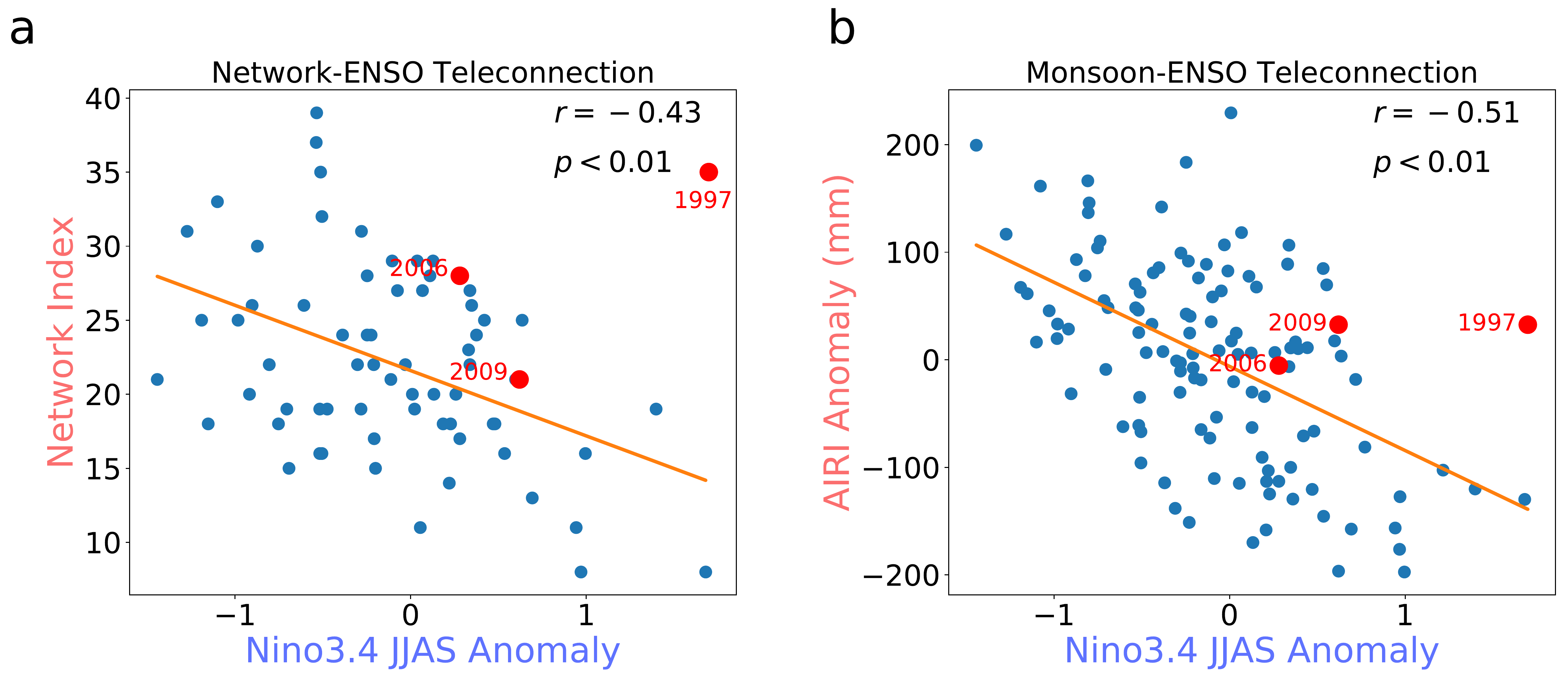}
\caption{\label{Fig6} 
{\bf The possible mechanism.} (a) The Network delayed  ENSO teleconnection. Plot of our network predictor and summer (JJAS) Ni\~{n}o3.4 SST anomaly (from 1950 to 2018). (b) The Monsoon-ENSO teleconnection.  Plot of standardized all-India summer monsoon rainfall and  Ni\~{n}o3.4 SST anomaly (from 1901 to 2018). The three red points represent the exceptional years 1997, 2006, 2009 detected by network approach [see Fig. S5]. The network index is calculated  based on the previous calendar year's data.}
\end{centering}
\end{figure}

The variability of the ISMR is  linked to the Pacific ENSO, and in particular some severe droughts over India are associated with \el~events \cite{webster_monsoon_1992,kumar_weakening_1999,rajeevan_nino-indian_2007}. Almost all the statistical seasonal prediction
models of ISMR rely heavily on the   magnitude change in various ENSO indices \cite{kumar_epochal_1999}. Numerical GCMs, which are used for seasonal rainfall prediction,  are also based on the SST specified in the Pacific \cite{sperber_interannual_1996}. 
It was found that the SST index over the central Pacific (Ni\~{n}o3.4) is a better indicator for the association between \el~and the ISMR than the combined Ni\~{n}o index derived from Ni\~{n}o 3 and Trans Ni\~{n}o Index \cite{rajeevan_nino-indian_2007,kumar_unraveling_2006}.
Therefore, to reveal the ENSO-Monsoon teleconnection, here we utilize the  Ni\~{n}o3.4 Index [SST anomaly], and find that the AIRI anomaly and the JJAS Ni\~{n}o 3.4 SST anomaly index are negatively correlated, i.e., $r=-0.51$ and $p <0.01$ (Fig. \ref{Fig6}b). Moreover, we present the spatial  correlation maps between the Ni\~{n}o 3.4 SST anomaly index with both the JJAS SST (Fig. S6a), as well as the rainfall index (Fig. S6b) for all nodes. We find that the Ni\~{n}o 3.4 SST anomaly index is significantly anti-correlated
%
with the SST in the equatorial Indian ocean and also heavily influences the rainfall pattern over the Indian land region. 

\subsection{Indian Homogeneous Region Rainfall Index}

The first official seasonal monsoon forecast was issued by Blanford in 1886. After that numerous models were developed for the seasonal forecasting of the ISMR.
Most of them are focused only on the prediction of the AIRI. However, the Indian monsoon precipitation 
shows distinct regional disparities, as demonstrated in Fig.~S7 the time series of the observed rainfall for All India and four homogeneous regions, including `Northwest India', `Central India', `East \& Northeast India',`South Peninsula'. The correlations with each other as well as the AIRI are summarized in Fig. S8.  We find that not all homogeneous regions' rainfall are significantly correlated with the AIRI.
It is thus a great necessity to design a rainfall prediction scheme for Indian homogeneous regions.  Here we show that our network-based approach can be also applied to forecast the Indian homogeneous region rainfall. In the present study, we perform our forecasting in two specific homogeneous regions: (1) Central India Rainfall Index (CIRI), since in recent years, the agriculture in the states of Madhya Pradesh, Jharkhand, Chhattisgarh in central India has been growing rapidly, and  relies heavily on monsoon \cite{gulati2017making}; (2) East \& Northeast India Rainfall Index (EIRI), since EIRI has a poor correlation (weak negative) with the AIRI, see Fig. S8.

Figs.~S7 and S8 suggest that the CIRI and the AIRI are strongly positively correlated ($r>0.8$), which indicates that they might have the same physical mechanism. Therefore, we use the same network predictor to forecast the CIRI. The results are shown in Fig.~S9.  We find that our forecasting skill is also significant with $r=0.43$, and
$p <0.01$. In particular, our prediction of the CIRI for 2019 is 1143.61 mm, compared  to  IMD's  observation value 1262.8 mm.
We also examine the forecasting ability of the same predictor for the EIRI. Not surprisingly the forecasting skill becomes very poor, as shown in Fig. S10.
This motivates us to rethink and redevelop the  East \& Northeast India Rainfall prediction method.

Analogue to our aforementioned predictors for the AIRI and the CIRI, to forecast the EIRI, we construct new CNs by only connecting global nodes with strong negative link strengths, i.e., negative links with the top 5\% absolute values [see Methods, Eq. (6)].
%
Then the new predictor for each node $i$ is defined as,
\begin{equation}\label{main1}
P^{y}_{i} = \sum_{j} A^{y}_{i,j}\cdot \max(C^{y}_{i,j}( \tau))
\end{equation}
where $A^{y}_{i,j}$  =1 (if i and j is connected) or 0 (otherwise), is the adjacency matrix of a CN for the year $y$ \cite{fan2017network}, $C^{y}_{i,j}( \tau)$ is defined in Eqs.~(\ref{eq1}) and (\ref{eq2}). 
Similar to the analysis for the AIRI, we mine the best network predictors for the EIRI in the training period, 1950-1989 (see Fig.~S11 a and b).
The optimized network predictor is given from the key node which is located in China (yellow box in Fig.~S11a) [(37.5\degree N, 85\degree E)].  Our forecasting skill based on the optimized network predictor  is also significant with $r=0.46$, and
$p <0.01$ [Fig.~S11c].
In addition, we predicted the EIRI for 2019 to be 1224.03 mm, compared  to  IMD's  observation value 1240.7 mm [1\% error]. 
%
 The potential physical mechanism for this new network predictors of the EIRI need further analysis based on climate
models and observation data.

In order to  measure the accuracy  of our forecasting skill, we also present the predicted  AIRI, CIRI, EIRI with 5\% (Fig. S12) 10\% (Fig. S13) error bars  versus their corresponding observed values. In particular, we find that for the AIRI predictions, 22 out of 30 years are within the 5\% error bars, and 28 out of 30 years are within the 10\% error bars.

\subsection{Climate Changes effect on Network-based Prediction}

How does climate change affect our forecasts? We examine the global temperature anomaly near-surface, at 1000 hPa and 850 hPa air pressure levels relative to the 1951-1980 average. We focus our discussion on the South Atlantic and Indian ocean areas with the aim to reveal the relation between global warming and the evolution of the in-degree index, which we use as a predictor. 

The change of the temperature anomaly for the last five years 2015-2019 relative to the 1951-1980 average, appears significant at 850 hPa air pressure level. There is a hotspots area in terms of  ``warmer than average'' in the mid-latitude South Atlantic that spreads around South Africa and extends into the mid-latitudes of the Indian Ocean (Fig.~\ref{Fig10_1_1}a). The difference in the temperature anomaly in this area is from 1.5 to 3 \degree C relative to the 1951-1980 average. It is about two times the magnitude compared with the 1000 hPa level where the effects of orography and diurnal variation blur the contrast (Fig. S14).  Remarkably,  the hotspots area concurs with the teleconnection path from the key node in South Atlantic through the chain of atmospheric circulations patterns to the Indian subcontinent (Fig. 3). Hence, the key node location belongs to the hotspots area.

Has the hotspots area appeared in the last five years 2015-2019 or it is a warming trend since 1948? We analyze the monthly temperature in the location of the key node from 1948 to 2019.  Using linear regression we find that even with the presence of decadal oscillations, there is a warming trend with a rate of  0.36 \degree C (Fig.~\ref{Fig10_1_1}b), i.e., for the last 70 years, the temperature in the key node area has risen $\sim$ 2.5 \degree C. 

Does the warming in the atmosphere agree with the evolution of the in-degree index? For comparison, we calculate the average temperature anomaly (using a five-year sliding window) for the key node for the last 40 years. Next, we  perform a similar analysis for the network predictor, the in-degree index, by using consecutive five-year periods. In both time series, superimposed in Fig.~\ref{Fig10_1_1}c, we uncover two increasing trends with $r\approx0.89$ ($r\approx0.75$) and $p$-values $<0.01$ for the in-degree index and the temperatures, respectively. We like to emphasize that the increase in the in-degree index is quite substantial, from 17 (for the period 1979-1983) to 32 (for the period 2014-2018). This increase means that the connectivity is enhancing in the considered region. We also find coherence between the two time-series. For example, both are low at the beginning of the 1990s and increase during the end of the 1990s. Thus, the warming in the key node area concurs with the rise of the in-degree index.
 
We further find that the warming trend affects not only the in-degree index but also the structure of the networks itself. With warming, we uncover an increasing number of long-distance links in the evolution of our network, see Fig.~\ref{Fig10_1}. Particularly, in the period 1979-1983, the strongest incoming links to the key node appear due to mid-latitude Westerlies and its interaction with the south polar jet stream (Fig.~\ref{Fig10_1}a). In the period 2014-2018, however, we uncover that there are more long cross-equatorial links appearing between Southwest Atlantic and North Asia-Pacific (Fig.~\ref{Fig10_1}b).
Our finding is consistent with observations: under anthropogenic global warming, the cross-equatorial flow became stronger \cite{hu_cross-equatorial_2018}. Here, we observe cross-equatorial links in the Indian Ocean, which penetrate through the South China Sea to North Asia-Pacific. Our result of the appearance of long cross-equatorial links is coherent with the observation of wind flow pattern in the Indian Ocean (Fig.~\ref{Fig10_1}b). 
 
How does the warming trend in the South Atlantic affect the prediction?  We analyze how network-based prediction skills are changing with the temperature increase. To quantify the forecasting skill, we use the deterministic Mean Square Skill Score (MSSS) and the correlation skill $r$ for two periods: the first 15 years 1989-2003 and the last 15 years 2005-2019. We find that both skill measures are improved under the warming trend for all three homogeneous regions in India. For example,  the MSSS and $r$ increase for each region as following:  all India rainfall index (AIRI) - from 0.078 to 0.384 and from 0.380 to 0.638;  for Central India rainfall index (CIRI) - from 0.156 to 0.332 and from 0.403 to 0.586; for the East India Rainfall Index (EIRI) - from -0.081 to 0.258 and from 0.151 to 0.533, respectively (Fig.~\ref{Fig10_1_1}d). These results strongly suggest that the prediction skills of AIRI, CIRI and EIRI are improved substantially for both MSSS and $r$ under global warming.

In summary, we find solid evidence that climate change affects our forecasts. Firstly, the key node location belongs to a hotspot area, with temperatures at 850hPa being 1.5 to 3 \degree C higher during 2015-2019 than the 1951-1980 average. Secondly, this significant warming trend substantially impacted this region and we observe this impact in the considerable increase of the in-degree index and the change in the structure of the network itself. Thus, the connectivity of the key node is enhanced resulting in the appearance of long cross-equatorial links. Thirdly, concurrently with the warming trend and the change of the network, we find that the prediction skill of our forecasting method for the Indian monsoon rainfall amount is also improving substantially.  Finally, our evidence clearly shows that climate change benefits our prediction.

\begin{figure}
\begin{centering}
\includegraphics[width=1.0\linewidth]{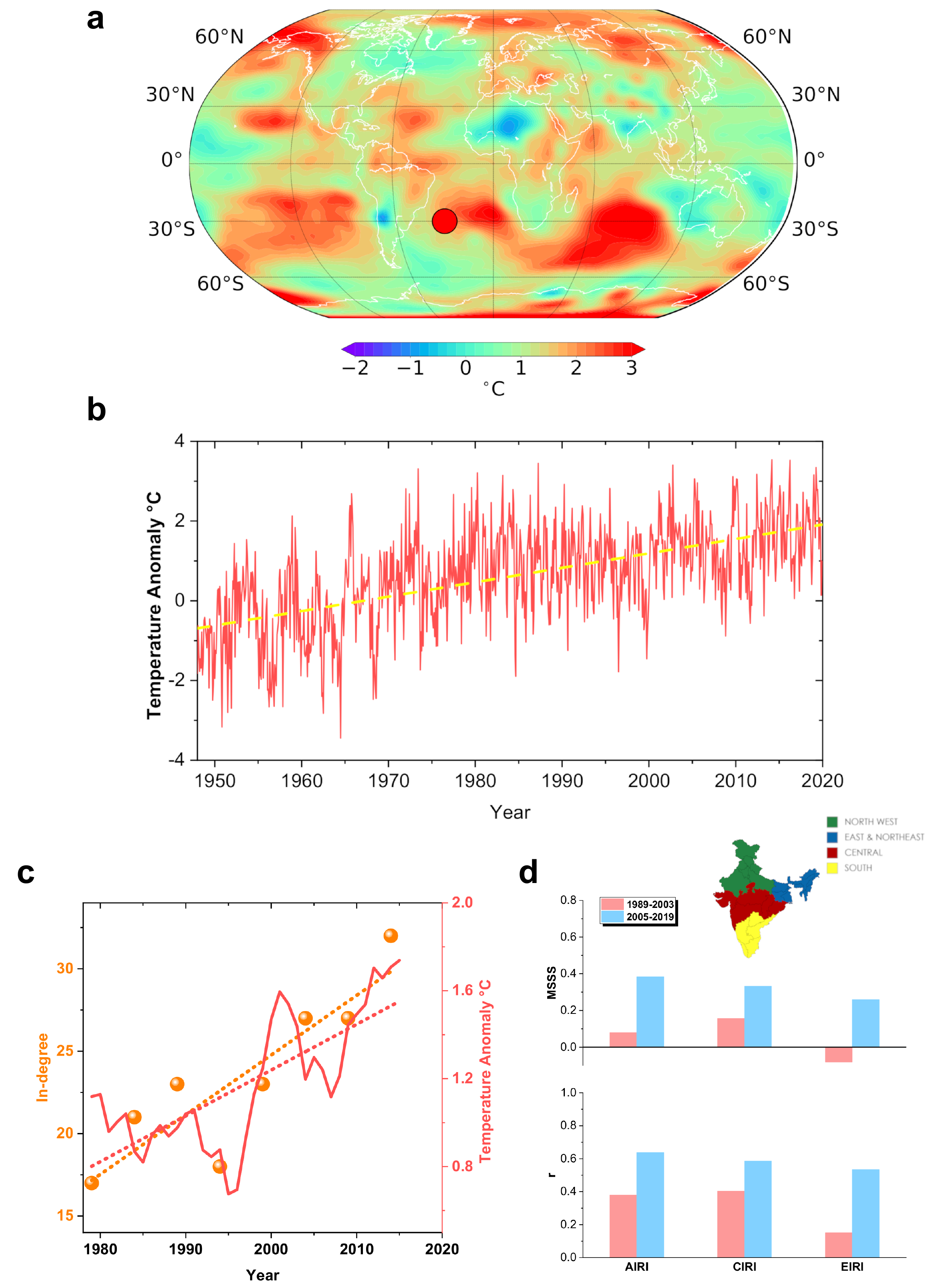}
\caption{\label{Fig10_1_1} 
{\bf Climate Change Benefits for Forecasting of India Monsoon Rainfall.}}
\end{centering}
\end{figure}
\begin{figure}[t]
  \contcaption{(a) Global average temperature anomaly for the last five years 2015-2019 relative to 1951-1980 average. The key node (red point) is in or very close to the climate hot spots, where the impacts of climate change are both pronounced and well documented in the south Atlantic. (b) The monthly average temperature anomaly at the key node relative to 1951-1980 average. The dashed line is the best fitting line with $r>0.6$, and $0.36$ $^\circ$C increasing per decade. (c) The network predictor, in-degree index,  and the five years' average temperature anomaly at the key node as a function of time. The dashed orange (red) line is the best fit-line for the in-degree index (temperature anomaly) with $r\approx0.89$ ($r\approx0.75$) and $p$-values $<0.01$. (d) Above: the deterministic forecasts skill Mean Square Skill Score (MSSS) and Below: the correlation skill $r$ for the first 15 years 1989-2003 and the last 15 years  2005-2019 for the All India Rainfall Index (AIRI, left), the Central India Rainfall Index (CIRI, middle) and the East \& North East India Rainfall Index (EIRI, right), respectively.  Inset in (d) shows the meteorological subdivisions of India. Here we used the NCEP/NCAR reanalysis monthly  air temperature at 850 hPa air pressure level.}
\end{figure}

\begin{figure}
\begin{centering}
\includegraphics[width=1.0\linewidth]{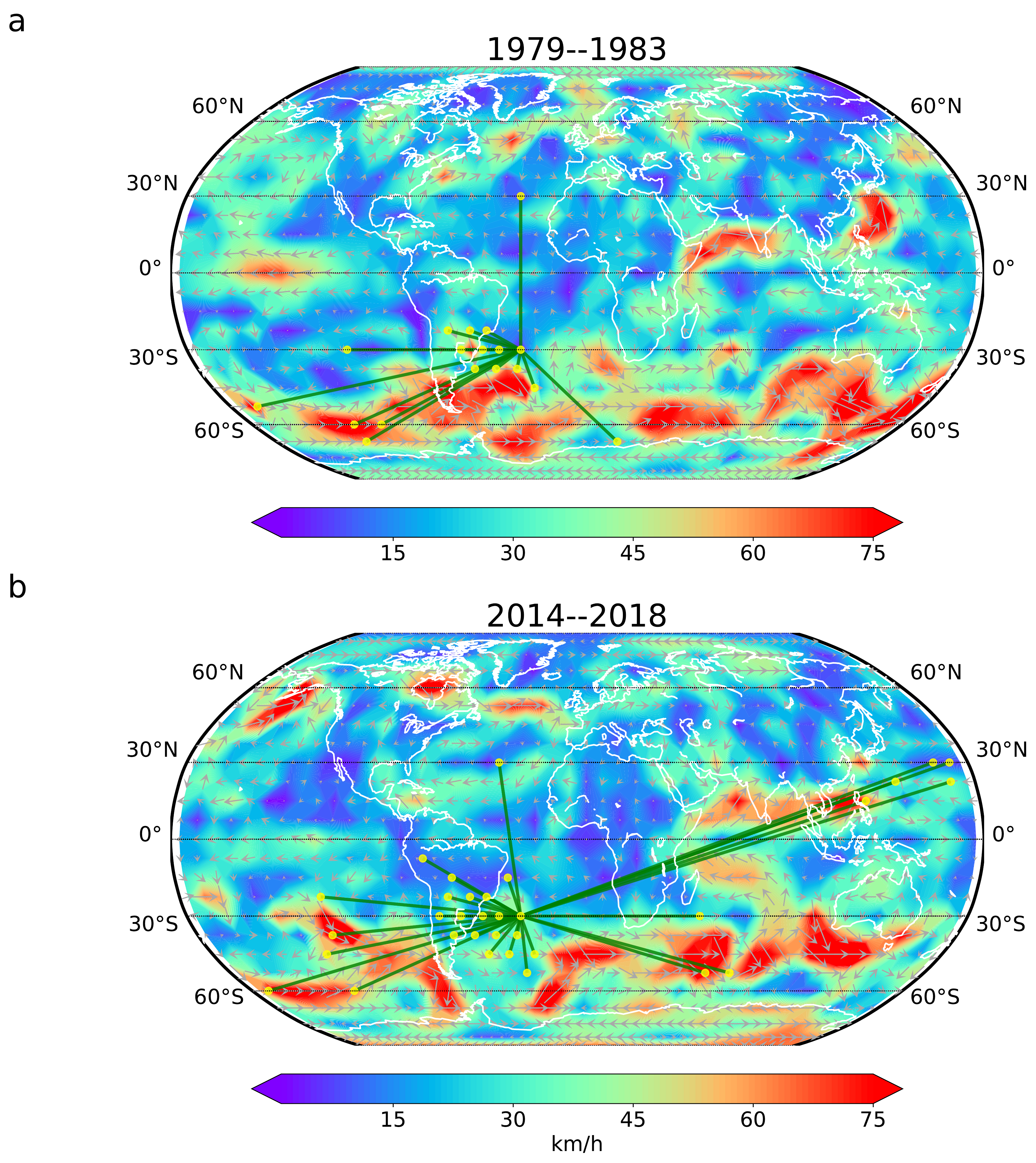}
\caption{\label{Fig10_1} 
{\bf Network structure examples for (a) 1979-1983 and (b) 2014-2018.} The background contains the information of the directions and magnitude of 850 hPa winds. Here we chose the wind data on 12th August 1982 for (a) and 2018 for (b) respectively.}
\end{centering}
\end{figure}


\section{Discussion}

Anthropogenic climate change   has  led  to  widespread  shrinking  of  the  cryosphere, rising  global mean-sea levels, increasing number of tropical cyclones,  and associated cascading impacts.  It also leads to significant socioeconomic ramifications in natural and human systems \cite{carleton_social_2016}. As  reported by the IPCC \cite{intergovernmental2018special}: abrupt warming in the Indian Ocean and extreme Indian Ocean Dipole events have largely altered the Asian monsoon, impacting the food and water security over these regions. The Indian summer monsoon circulation and rainfall exhibit a statistically significant weakening since the 1950's. This weakening has been hypothesised to be a response to the Indian Ocean basin-wide warming
\cite{mishra_prominent_2012,roxy_drying_2015} and also to increased aerosol emissions \cite{guo_local_2016}. An increase in extreme rainfall events occurred at the expense of weaker rainfall events \cite{goswami_increasing_2006} over the central Indian
region, and in many other areas \cite{krishnamurthy_relation_2009}.
The strongest effect of climate change on the monsoon is the increase in atmospheric moisture associated with
warming of the atmosphere, resulting in an increase in the total monsoon rainfall even when the strength of the monsoon
circulation weakens or does not change \cite{intergovernmental2018special}. CMIP5 models project an increase in mean precipitation of the Indian monsoon largely due to the increased moisture flux from ocean to land \cite{christensen2013climate}. These models also project an increase of the  interannual variability and extremes of the Indian monsoon. Particularly concerning is that 
%
%
%
 the AIRI predictions for both operational statistical models and GCMs have recently failed severely \cite{wang_rethinking_2015}. This is because the successes of seasonal forecasts of the ISMR in these models rely on the robustness of the ENSO-monsoon teleconnection, i.e.,  the inverse relationship between the ENSO and ISMR.
However, this teleconnection has weakened in recent decades \cite{kumar_weakening_1999}. For example, the official operational model made by IMD predicted an abnormally low precipitation (dry) for the monsoon season during 1997 which is the onset year of a very strong El Ni\~{n}o event.
Counterintuitively, it was found to be a false prediction and no drought occurred during that year \cite{kumar_unraveling_2006}. Even worse, it was also reported that extreme \el ~events are projected to likely increase in frequency in the 21st century \cite{cai_increasing_2014}, which indicates that the forecasting uncertainties of the ISMR by the conventional models may become higher.

Network theory allows predicting several climate phenomena such as \el~ events \cite{ludescher_very_2014,meng_forecasting_2018,meng2019complexity}, monsoon arrivals/retreats \cite{stolbova_tipping_2016}, 
extreme regional precipitation patterns \cite{boers_prediction_2014} or anomalous polar vortex dynamics \cite{kretschmer_early_2017} much earlier than conventional schemes that rely on direct simulations of fluid dynamics through sophisticated models. Most importantly, network analysis can serve both as a practical generator of early-warning approaches as well as a scientific inspiration for mapping out the topological features that are related to  physical mechanisms of the dynamic systems. 


To  fill the gap
in the ISMR prediction, 
 here, we have developed a network-based framework.  Thus a long-term (5-months ahead) seasonal forecasting for the AIRI, CIRI and EIRI is achieved. We find that the introduced prediction method yields a forecast skill of $\sim 0.5$, which is comparable to the best statistical and dynamical forecast models, however, with quite a short (i.e., within one month) lead-time.  We propose a possible physical mechanism underlying our predictor and associate it with the network--delayed--ENSO and ENSO--monsoon teleconnections.
Moreover, we find solid evidence that climate change affects our forecasts. The considerable warming trend in the last 70 years -the 850 hPa temperature in the key node area has risen 2.5 \degree C- substantially impacted the key node region. We observe an increase of the in-degree index and the change in the structure of the network itself. Concurrently with these changes, we find that the prediction skill of our forecasting method for the Indian monsoon rainfall amount, AIRI, CIRI and EIRI, is improving substantially. 
The proposed methodology offers the following advantages: it forecasts the amount of rainfall 5-months in advance and offers the opportunity for regional forecasting. 
Under climate change, as floods and droughts will become more frequent, long range and reliable forecasts help to prepare national and regional resources for disaster management and to strengthen resilience against disruptive weather phenomena.

Finally, we forecast here the future 2020 monsoon i.e., the AIRI, CIRI and EIRI to be 842.85 mm (-5\% below average), 854.07 mm (-12\%) and 1245.54 mm (-13\%). These results indicate that compared to the normal rate, 
the predicted deficit of the monsoon rainfall might be caused by the oncoming moderate or strong \el~ at the end of 2020 \cite{ludescher2019very,meng2019complexity}.

%

If our approach can be confirmed by further empirical and theoretical studies, a new framework in the
scientific field of forecasting may be initialized. There would be far-reaching implications for
operational long term forecasts deployed around the world.
Our results based on network theory provide a new perspective to set new research in climate prediction
 and help stakeholders to improve  disaster-prevention agendas.

\section{Data and Method}

\subsection{Data sets used in this study}
(i) All-India rainfall data, and the four homogeneous regions including `Northwest India', `Central India', `East \& Northeast India', `South Peninsula' rainfall data   provided by IMD (1901-2018), \url{https://data.gov.in/catalog}.  (ii) Oceanic Ni\~{n}o Index  and Ni\~{n}o 3.4 Index from National Climatic Data Center (NOAA), \url{https://origin.cpc.ncep.noaa.gov/products/analysis_monitoring/ensostuff/ONI_v5.php}. (iii) NCEP/NCAR reanalysis monthly means surface or near the surface (.995 sigma level) air temperature  ~\cite{kalnay_ncepncar_1996} (1949--2019), \url{https://www.esrl.noaa.gov/psd/data/gridded/data.ncep.reanalysis.derived.surface.html}.  (iv)  NCEP/NCAR reanalysis daily surface or near the surface (.995 sigma level) air temperature ~\cite{kalnay_ncepncar_1996} (1948--2019), \url{https://www.esrl.noaa.gov/psd/data/gridded/data.ncep.reanalysis.surface.html}. It has a spatial (zonal and meridional) resolution of $2.5^\circ \times 2.5^\circ$, resulting in 10512 grid points. (v) Global Precipitation Climatology Project monthly precipitation dataset (v2.3) from 1979-present combines observations and satellite precipitation data into $2.5^\circ \times 2.5^\circ$ global grids \cite{adler_version-2_2003}, \url{https://www.esrl.noaa.gov/psd/data/gridded/data.gpcp.html}. (vi) For SST validation, we use  NOAA's Extended Reconstructed SST (v5) at
$2^\circ$ spatial resolution for the period 1854--2018 \cite{huang_extended_2017}, \url{https://www.esrl.noaa.gov/psd/data/gridded/data.noaa.ersst.v5.html}.
(vii) NCEP/NCAR reanalysis daily 850 hPa air temperature and data, \url{https://www.esrl.noaa.gov/psd/data/gridded/data.ncep.reanalysis.pressure.html}.

For each node $i$ (i.e., longitude-latitude grid point), we calculate the daily atmospheric temperature
anomalies $T_i(t)$ (actual temperature value minus the climatological average and divided by the climatological standard deviation) for each calendar day. For the calculation of the climatological average and standard deviation, only past data up to the prediction date have been used. For simplicity leap days were excluded.  We used the first 41 years of data (1948-1988) as a training set to calculate the first average value and start the prediction from 1989.

\subsection{Climate Network}
\subsubsection{Nodes}
In the current work we select 726 nodes \cite{gozolchiani_emergence_2011} (shown in small dots in Fig. 1a), such that the globe is covered approximately homogeneously.
\subsubsection{Links}

For obtaining the time evolution of the strengths of the links between each pair of nodes $i$ and $j$, we follow~\cite{yamasaki_climate_2008} and compute, for each year $y$, the time-delayed, cross-correlation function
\begin{equation}\label{eq1}
C^{y}_{i,j}(-\tau)=\frac{\langle T_i^{y}(t) T_j^{y}(t-\tau) \rangle-\langle T_i^{y}(t)\rangle \langle T_j^{y}(t-\tau) \rangle }{\sqrt{\langle (T_i^{y}(t)-\langle T_i^{y}(t)\rangle)^2\rangle}\cdot\sqrt{\langle (T_j^{y}(t-\tau)-\langle T_j^{y}(t-\tau)\rangle)^2\rangle}},
\end{equation}
and
\begin{equation}\label{eq2}
C^{y}_{i,j}(\tau)=\frac{\langle T_i^{y}(t-\tau) T_j^{y}(t) \rangle-\langle T_i^{y}(t-\tau)\rangle \langle T_j^{y}(t) \rangle }{\sqrt{\langle (T_i^{y}(t-\tau)-\langle T_i^{y}(t-\tau)\rangle)^2\rangle}\cdot\sqrt{\langle (T_j^{y}(t)-\langle T_j^{y}(t)\rangle)^2\rangle}},
\end{equation}
where the brackets denote an average over the past 365 days, according to
\begin{equation}
\langle f(t) \rangle=\frac{1}{365}\sum_{a=1}^{365}f(t-a).
\label{eq3}
\end{equation}
 We consider, for the daily datasets, time lags of $\tau \in [0, 200]$ days, where a reliable estimate of the background noise level can be guaranteed (the appropriate time lag is discussed in \cite{guez_influence_2014}).

Next, similar to ~\cite{fan2018climate}, we define the positive and negative strengths of the link between each pair of nodes (e.g., nodes i and j) in the network as below
\begin{equation}
W^{+,y}_{i,j} = \frac{\max(C^{y}_{i,j}) - {\rm mean}(C^{y}_{i,j})}{{\rm std}(C^{y}_{i,j})},
\label{eq4}
\end{equation}
and 
\begin{equation}
W^{-,y}_{i,j} = \frac{\min(C^{y}_{i,j}) - {\rm mean}(C^{y}_{i,j})}{{\rm std}(C^{y}_{i,j})},
\label{eq5}
\end{equation}
where ``max'', ``min'', ``mean'' and ``std'' are the maximum, minimum,  mean and standard deviations of the cross-correlation function.

\subsubsection{In-degree}
The links are sorted in decreasing (increasing) order of strength and then added one by one according to decreasing (increasing) strength $W^{+} (W^{-})$ [see Eq.
(5) and (6)]; i.e., we first choose the most important link with the highest weight, then the second strongest link, and so on. In the current study, we consider the top 5\%  (95\% confidence level) as significant links. 

Meanwhile, we identify the value of the highest peak in the cross-correlation function and denote the corresponding time lag of this peak as $\theta^{y}_{i,j}$ \cite{fan2017network}.  The sign of $\theta^{y}_{i,j}$ indicates the direction of each link; i.e., when the time lag is positive ($\theta^{y}_{i,j} > 0$), the direction of the link is from $i$ to $j$. For each node $i$, we construct its in-degree time series $K^{y}_{i}$ by counting the number of links  to this node $i$ for each calendar year $y$. 

\subsection{The time of the forecast}
Using our network-based approach, a forecast of the ensuing AIRI, CIRI and EIRI can be made based on the previous calendar year's data, while  the operational
forecast starts from 1 June.
\subsection{Null-model}
To further demonstrate that all added links in our CN are significant, we consider the Null-model. 
In the Null-model, we compute the strength of links of surrogate time series (reshuffled regional data by days). We then repeat this process 100 times and obtain the Null-model links' distributions.
\subsection{Mean square skill score}
The Mean Square Skill Score \cite{murphy_skill_1988} (MSSS) is used to measure the deterministic forecasts skill.
The MSSS is defined as follows:
\begin{equation}
MSSS  = 1 - \frac{MSE_{forecast}}{MSE_{c}}.
\label{eq7}
\end{equation}
Where the mean squared error of the forecasts is:
\begin{equation}
MSE_{forecast}  = \frac{1}{n} \sum_{i=1}^{n} (f_{i} - x_{i})^2,
\label{eq8}
\end{equation}
where $x$ and $f$ denote time series of observations and forecasts. 
The MSE for climatology is given by:
\begin{equation}
MSE_{c}  = \frac{1}{n} \sum_{i=1}^{n} (x_{i} - \<x\>)^2,
\label{eq9}
\end{equation}

\section*{Acknowledgements}
We acknowledge S. Havlin, Y. Ashkenazy, Y. Zhang and N. Yuan for their helpful suggestions.
We thank the ``East Africa Peru India Climate Capacities --- EPICC'' project, which is part of the International Climate Initiative (IKI). The Federal Ministry for the Environment, Nature Conservation and Nuclear Safety (BMU) supports this initiative on the basis of a decision adopted by the German Bundestag. 
E. S. acknowledges the support from the RFBR (No. 20-07-01071). L. Z. acknowledges the support from the PFRSF (PF01001160) and the NSFC (No. 11901492).

\section*{Contributions} All authors designed the research, analyzed data, discussed results, and contributed to writing the manuscript.

\section*{Additional information}
Supplementary Material is available in the online version of the paper. 
\section*{Data and Code availability}
The data and codes that support the findings of this study are available from the corresponding author upon reasonable request.

\section*{Competing financial interests}
The authors declare no competing financial interests.

\bibliographystyle{naturemag}
\bibliography{MyLibrary}

\end{document}